\documentclass[final,1p,times,twocolumn]{elsarticle}

\usepackage{amssymb}
\usepackage{amsmath}
\usepackage{amsthm}

\usepackage{lineno}
\usepackage{booktabs} 
\usepackage{float}

\usepackage{stfloats}

\begin{document}

\begin{frontmatter}



\title{Automated Rib Fracture Detection of Postmortem Computed Tomography Images Using Machine Learning Techniques}

\author[label 1]{Samuel Gunz}
\author[label 1]{Svenja Erne}
\author[label 2]{Eric J. Rawdon}
\author[label 1]{Garyfalia Ampanozi}
\author[label 1]{Till Sieberth}
\author[label 1]{Raffael Affolter}
\author[label 1]{Lars C. Ebert}
\author[label 1]{Akos Dobay}
\address[label 1]{Zurich Institute of Forensic Medicine, University of Zurich, 8057 Zurich}
\address[label 2]{Department of Mathematics, University of St. Thomas, St. Paul, Minnesota, 55105-1079}

\begin{abstract}
Imaging techniques is widely used for medical diagnostics. This leads in some cases to a real bottleneck when there is a lack of medical practitioners and the images have to be manually processed. In such a situation there is a need to reduce the amount of manual work by automating part of the analysis. In this article, we investigate the potential of a machine learning algorithm for medical image processing by computing a topological invariant classifier. First, we select retrospectively from our database of postmortem computed tomography images of rib fractures. The images are prepared by applying a rib unfolding tool that flattens the rib cage to form a two-dimensional projection. We compare the results of our analysis with two independent convolutional neural network models. In the case of the neural network model, we obtain an $F_{1}$ Score of 0.73. To access the performance of our classifier, we compute the relative proportion of images that were not shared between the two classes. We obtain a precision of 0.60 for the images with rib fractures.
\end{abstract}


\end{frontmatter}


\section{Introduction}

Deep learning commonly describes a type of machine learning technique used for image classification. Although the technique has existed since the 80s \cite{lecun1985,lecun1989,lecun2015} it is only in the last decade that deep learning has attracted some attention. The interest in deep learning can be explained by the constant evolution of graphics processing units (GPUs) and their ability not only to manipulate computer graphics, but also for general purposes, and this at very competitive prices. Today, most of the deep learning open source frameworks take advantage of GPU acceleration for image analysis. As a result of this evolution, and the amount of data collected through online service providers, deep learning is widely used to improve users’ experience by classifying customer profiles or medical diagnosis based on various imaging techniques such as computed tomography (CT), magnetic resonance imaging (MRI), positron emission tomography (PET), and X-ray imaging. Furthermore, deep learning has been implemented in automotive solutions to produce self-driving cars and to automate traffic control. However, classification problems are common in fields other than image processing. One domain where mathematicians are dealing with the classification of objects having variable size, shape, and orientation is topology.  The mathematical field of topology investigates properties of geometric forms that remain invariant under certain transformations such as bending or stretching. An example of a topological object is a closed curve embedded in three-dimensional Euclidean space. Such a closed curve is also the basis for the mathematical definition of a knot. Hence, the words ``closed curve'' and `knot'' are interchangeable, and studies dedicated solely to closed curves embedded in three-dimensional Euclidean space constitute a subfield of topology called knot theory. To distinguish two knots, a two-dimensional projection onto a Euclidian plane is produced. If the curve crosses itself in the projection, a convention is applied at the crossing to indicate which branch lies under by splitting the projected line into two segments while keeping the other branch straight between the two segments (see the supplemental section for more details). However, some projected directions may produce a situation where this convention cannot be applied. Figure 1 in the supplemental information shows the degenerate cases. A polynomial knot invariant is a polynomial having one or two variables. The polynomial remains unchanged regardless of the curve orientation during its projection. A very first variant of a polynomial knot invariant was proposed by J. W. Alexander in 1928 \cite{alexander1928}. More elaborate polynomials have been proposed since the 1980s. Here, we used the HOMFLY polynomial, a generalized Jones polynomial using two variables \cite{jones1985, freyd1985}.

\begin{figure*}
\includegraphics[keepaspectratio = true, height=2.1 in]{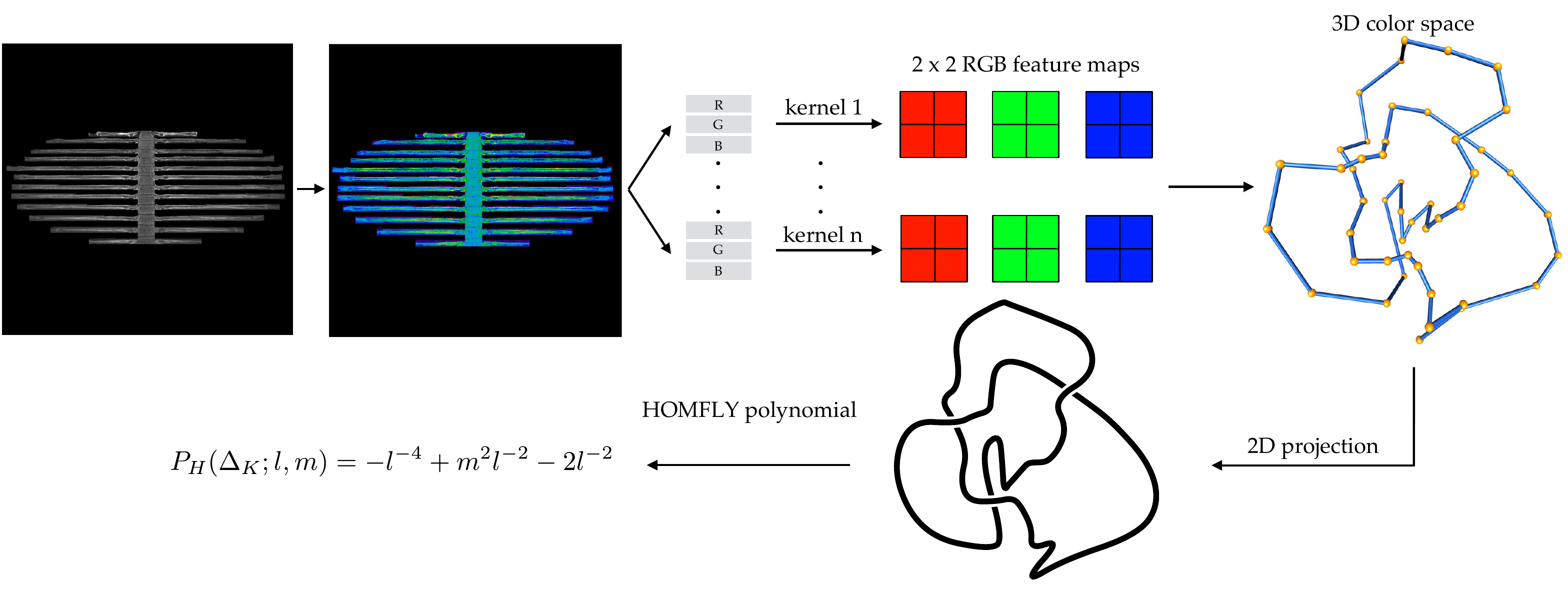}
\caption{Flowchart depicting the calculation of a polynomial knot invariant starting from a CT image. The first step consists of finding a projection of the unfolded rib cage with the most informative view while rotating the ribs. The greyscale CT image of the projection is subsequently transformed into RGB using the colormap function \texttt{nipy\_spectral} of the \texttt{Python} module \texttt{Matplotlib}. For each channel of a given input image, we apply separately one of the 13 kernels in multiple rounds of convolution and pooling to obtain a $2 \times 2$  feature map. From the feature map a set of four coordinates $(x_R, x_G, x_B)$ are extracted amounting to a total of 52 coordinates when combining all 13 kernels. These coordinates are joined to form a single piecewise-linear curve in a 3D RGB color space. The curve is then projected to determine the corresponding HOMFLY polynomial.}
\label{fig1}
\end{figure*}

Progress has already been reached in computed tomography (CT) based volume measurements in cases of pericardial effusion \cite{ebert2012}, automated segmentation for CT volumes of livers \cite{umetsu2014} and automated tumor volumetry in brain tumors \cite{gaonkar2015}. However, clinical radiologists need to view an image every 3 seconds on a regular working day. With increasing image complexity due to the large amount of data, the effectiveness of diagnosis is impaired \cite{collado2018, andriole2011}. As rib fractures are a common and important finding in forensic case assessment as a result of blunt force trauma or chest compression due to resuscitation attempts, automated rib fracture detection on CT images can greatly facilitate the work of forensic radiologists \cite{schulze2013}. Especially in cases of incomplete rib fractures, autopsy fails to detect the fractures and the value of postmortem computed tomography (PMCT) has increased \cite{schulze2013}. Among those, the so-called “buckle rib fractures” as a subgroup of incomplete fractures with symmetrical distribution and continuity along the mid-clavicular lines are commonly observed after cardiopulmonary resuscitation \cite{yang2013}. Hence, the correct detection and classification of rib fractures by a machine learning algorithm can be of forensic relevance, but also for clinical radiologists in general \cite{cattaneo2006,glemser2017}. Rib unfolding is a tool representing a single-in-plane image reformation of the rib cage by using a reconstruction algorithm. A spider-like image is produced with the vertebral column in the middle and on each side of the 12 ribs, which can be rotated around their transversal axis for interpretation.
In this paper, we are interested in implementing a machine learning technique by convolving images to produce a topological invariant for image classification. We tested the method on medical images, more specifically on rib fractures of PMCT images. We describe here the various steps in implementing our machine learning technique and measuring its performance. Figure \ref{fig1} gives an overview of the whole architecture.

\section{Materials and Methods}

\subsection{Ethics}
Data use for this study conforms with the Swiss laws and ethical standards as approved by the Ethics Committee (written approval, KEK ZH-Nr. 15-0686).  Cases were assigned identification numbers and were fully anonymized for image retrieval.

\subsection{Case selection}
A total of 340 consecutive postmortem cases were retrospectively selected from July 2017 up to April 2018 from our database. We excluded cases with signs of advanced decomposition (based on the radiological alteration-index (RA-Index) defined by Egger \textit{et al.} \cite{egger2012}), organ explantation and severe trauma with extensive damage to the corps such as amputation or exenteration,  cases with deviating scanning protocol or cases without PMCT, cases in which the rib fracture was not visible in the rib unfolding tool or in which the rib defect was in the cartilaginous part of the rib, as well as ongoing cases. One case was excluded due to accessory ribs with a gunshot defect in one of these accessory ribs which could not be displayed in the rib unfolding tool. In the end, 195 remaining cases were included. All 195 included cases underwent whole body native PMCT and a standard forensic autopsy was performed thereafter. Of these,  85 had only new rib fractures, 84 had no rib fractures, and 26 had old fractures with or without new fractures. Every kind of rib fracture was included, namely complete and incomplete fracture, independent of their location.

\subsection{Image treatment prior to classification}
The rib fracture images were extracted from volumetric CT data (see Figure 2 for more details in the supplementary materials) using Syngo.via rib unfolding tool (Siemens Healthineers GmbH, Erlangen, Germany) with standard window settings (center 450, width 1500). The value of a single voxel on PMCT images correspond to the X-ray absorption of the tissue, given in Hounsfield Units (HU). Each pixel has a value depth of 12 bits with values starting at -1000 (X-ray density of air), zero being the X-ray density of water, up to 3096 HU. The extracted images were encoded in a value depth of 8 bits.

\subsection{Data mining}
To differentiate between cases with and without rib fractures, we relied upon imaging findings. If there were imaging findings and they were visible on the reconstructed ribs as identified by a medical student one year before graduation and a certified board forensic pathologist with 13 years of forensic pathology experience and 9 years of forensic imaging experience, then it was classified as a fracture. The data were retrieved from the multi-modality image reading Software Syngo.via. In some cases the reconstruction was insufficient, e.g. following mis-segmentation.  Nevertheless we included these cases since these artefacts looked different when compared to rib fractures. Ringl \textit{et al.} \cite{ringl2015} also described this impediment. However, in their study, the authors tried to correct the segmentation with more than 4 affected ribs per image. For each case, we exported 3 images with random axial rib rotations and all graphics and lines hidden. The pictures were stored as portable graphic (PNG) documents. In the end, a total of $256$ images with new rib fractures, $253$ images without rib fractures, and $79$ images with old rib fractures with or without new fractures were collected.

\subsection{Convolution and dimensionality reduction}
For each image, we generated a closed curve by joining all the coordinates in three-dimensional Euclidean space. The coordinates were produced by flattening the $2 \times 2$  feature map of the final pooling to obtain four single pixel values per channel. The four single pixel values per channel were combined to produce four coordinates per filter. A total of 13 kernel functions out of the 28 listed in Table \ref{tab:filter} were chosen to generate a final set of 52 coordinates per image. This choice was motivated by the need to reduce the number of crossings when projecting the curve. The corresponding polynomial knot invariant was computed and the closed curve was classified based on its polynomial expression (see the HOMFLY polynomial section for more details). For each class the precision was calculated by computing how many polynomials ended in a different class. In this study we used a binary classifier system.

For this study a single image presenting all ribs was needed. The automatic rib unfolding tool was used. Dankerl \textit{et al.} \cite{dankerl2017} and Ringl \textit{et al.}  \cite{ringl2015} demonstrated improved diagnostic performance and minimized reading time when using the single-in-plane image reformation of the rib cage compared to the multiplanar reformats in transverse, coronal, and sagittal orientations, encouraging the use of this tool for this study.

\begin{figure}
\begin{center}
\includegraphics[width=5.3 in]{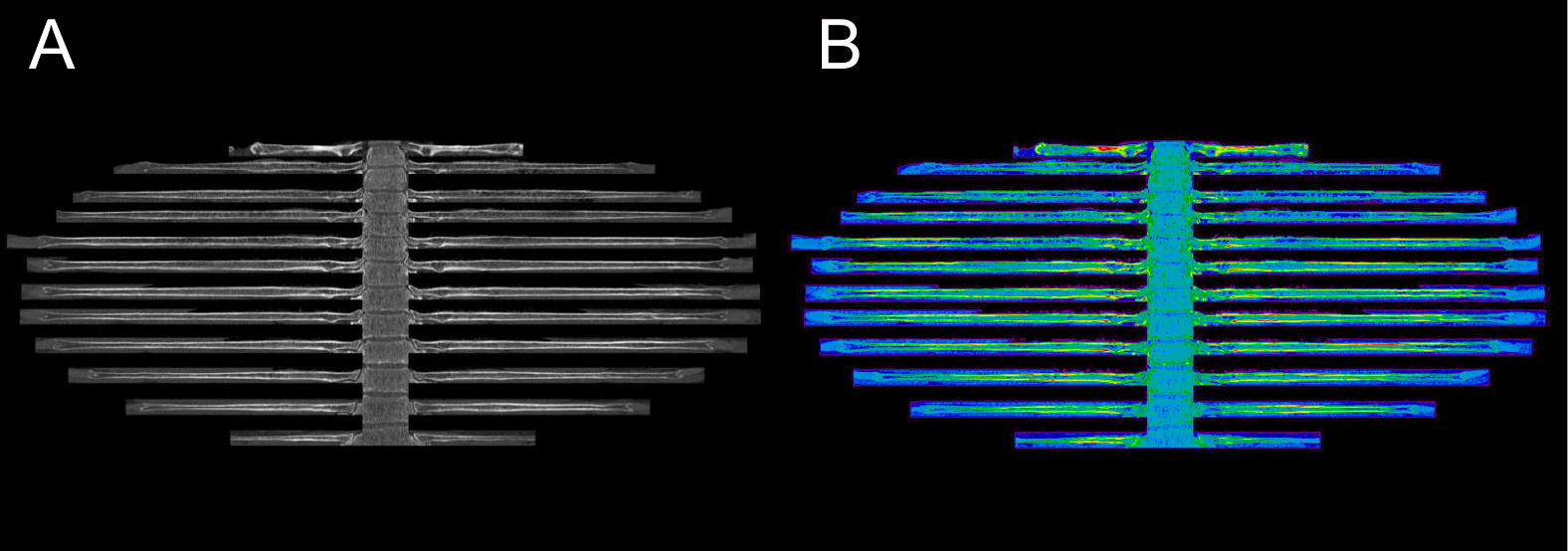}
\caption{Colormap conversion applied to transversal CT images of a human rib using the \texttt{nipy\_spectral} function of the Python module \texttt{Matplotlib}. (A) Original grayscale image showing a rib cage. (B) The same image after applying the color mapping function.}
\label{fig2}
\end{center}
\end{figure}

We implemented the design of our machine learning technique using standard libraries and the scikit-learn (scikit-learn.org) module for machine learning in \texttt{Python}. We defined the three RGB color channels of an image as the three axes of the color space. Color images are most commonly displayed with use of the Red, Green, and Blue (RGB) color model. Therefore, RGB images can be represented as a 3D array where each dimension corresponds to one color channel. If it is a grayscale PMCT image, a pseudo-color image can be derived from it by mapping each pixel value to a color. This can be achieved for instance by using the function colormap of the \texttt{Python} module \texttt{Matplotlib} \cite{hunter2007}. The colormap function of the  \texttt{Python} module uses the CIELab convention. In CIELab, the color space is represented by lightness (L), red-green (a), and yellow-blue (b). The optimal colormap depends on the type of images. We used the colormap \texttt{nipy\_spectral} due its contrasting range of colors. Figure \ref{fig2} shows an example of such a colormap conversion on a PMCT image of the rib cage with and without fracture. Since the input images were all in grayscale, but already encoded in RGB format, the values of each channel were identical. Therefore, only one channel was extracted to create the RGB images. Next, the channels of the pseudo-color images were separated to produce three different images, one for each channel. A total of seven different types of kernel were tested to generate the feature maps: blur, sharpen, edge enhancement, edge detection, line detection, emboss, and Sobel. Table \ref{tab:filter} gives a summary of the number of kernels for each type.
We used a $950 \times 950$ pixel resolution when processing the feature maps with the polynomial knot invariant and a $60 \times 60$ pixel resolution when using the custom built model in \texttt{Keras}. To produce a set of coordinates in three-dimensional color space we convolved each channel of the input images with a $3 \times 3$ kernel function (see Table \ref{tab:filter} for more details) using stride 1 followed by a downsampling using the softmax function of the \texttt{Python} module \texttt{NumPy}. This operation produced separate feature maps corresponding to each of the three RGB channels. We repeated this operation until we obtained a $2 \times 2$ convolved image for each channel. The set of coordinates $(x_R, x_G, x_B)$ were derived using the corresponding RGB channel of the four different values in the final feature map. To be able to calculate a knot invariant for the different images, we projected the three-dimensional piecewise linear curve obtained by joining all the 52 coordinates of the feature maps in the three-dimensional color space (see Figure \ref{fig1} for more details).

\subsection{Kernel analysis and number of convolution cycles}
\begin{figure}
\begin{center}
\includegraphics[width=5.3 in]{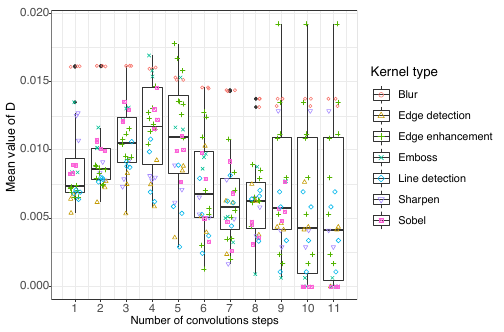}
\caption{Box plot of the optimal number of convolution steps for each kernel listed in Table \ref{tab:filter}.}
\label{fig3}
\end{center}
\end{figure}

To test the efficacy of each kernel on our dataset the coordinates related to the same filter were averaged by computing the centroid of each feature map. The number of convolution cycles was determined to be 4 for the current dataset. This value was obtained by maximizing the following quantity

\begin{equation}
 D = \frac{|| v_c^1 - v_c^2 || }{\sum_{i=0}^{N} || v_i^1 - v_c^1 || + \sum_{j=0}^{M} || v_j^2 - v_c^2 ||}
\end{equation}

\noindent
where $D$ is the relative fraction of the Euclidian distance between the centroid $v_c^1$ and  $v_c^2$ of two distinct filters divided by the spread of the two clusters. Figure \ref{fig3} shows the distribution of $D$ for the $28$ kernels listed in Table \ref{tab:filter}.

\begin{table}
\begin{center}
  \caption{Kernel functions: a total of $28$ kernels were tested.}
  \label{tab:filter}
  \begin{tabular}{lr}
  \\
    \toprule
    Kernel type&Number\\
    \midrule
    Blur & 3\\
    Sharpen & 3 \\
   Edge enhancement & 8\\
   Edge detection & 3\\
    Emboss & 3\\
    Sobel & 4\\
   Line detection & 4\\
  \bottomrule
\end{tabular}
\end{center}
\end{table}

\subsection{HOMFLY polynomial}
We used the HOMFLY polynomial as polynomial knot invariant \cite{freyd1985}. The HOMLFY polynomial can distinguish a large number of knots and it is sensitive to chirality: a knot and its mirror image tipically have different polynomial expressions. It is a generalised Jones polynomial with 2 variables. To compute the HOMFLY polynomial we used a program written by Bruce Ewing and Kenneth Millett \cite{ewing1997}. The coordinates in the three-dimensional color space were joined to create a closed piecewise linear curve which then served as proxy to compute the polynomial (see Figure \ref{fig1} for more details). For each image, a set of 52 coordinates were generated through the 13 kernels and the $2 \times 2$  feature maps. In order to assess the classification capabilities of the new approach, we used custom-made \texttt{Python} scripts based on \texttt{Tensorflow} version 1.8 \cite{abadi2016} and MobileNets version 1.0 \cite{howard2017}. The MobileNet architecture has been developed by researchers at Google to reduce computing time for embedded vision applications. The network parameters have been optimized on ImageNet classification. Our \texttt{Python} scripts retrain the output layer of the MobileNet architecture using the actual labels. We also developed a second model using the high-level neural network APIs \texttt{Keras} (https://keras.io). Using \texttt{Keras} APIs, we designed from scratch a fully connected neural network comprising three convolutional and two dense layers.

\subsection{Training and validation}
For training the models in \texttt{TensorFlow} we selected $127$ CT images with no fractures and $129$ with fractures (see Materials and Methods for more details). For validation we had $126$ with no fractures and $128$ with fractures. In the first case, we used the MobileNet architecture with a learning rate of 0.05 and 5,000 training steps. In the second case, we used a custom-made fully connected neural network consisting of three convolution layers and two dense or fully connected layers using \texttt{Keras} APIs. Here we used a dropout of 0.25 after the last pooling and another dropout of 0.5 before the dense output layer, a learning rate of 0.001 and we trained our model for 50 epochs. Recall, precision and $F_1$ Score was computed using the confusion matrix. In case of the polynomial knot invariant the precision was modified to account for the relative proportion of knot types that were not shared between the two classes, i.e.

\begin{equation}
\cfrac{K_1 - K_{12}}{K_1}
\end{equation}
 
\noindent
where $K_1$is the total number of knot types in one class and $K_{12}$ is the total number of knot types in both classes.

\section{Results}
For the first part using MobileNet, we achieved an overall recall of 0.68 and a precision of 0.76. The $F_1$ Score was 0.72. Using the custom-built model consisting of three convolution and two dense layers, we achieved a recall value of 0.68 and a precision of 0.79 leading to an $F_1$ Score of 0.73. Finally, we used the HOMFLY polynomial knot to classify the rib fracture images. The polynomials were matched to their corresponding knot type using the standard Alexander-Briggs \cite{briggs1927} notation, and the Hoste, Thistlethwaite, and Weeks notation for more complex knots \cite{hoste1998} . Due to the large number of possible knot types with 52 edges, even if the HOMFLY polynomial can be calculated from a projection it is not straightforward to identify its corresponding knot type unless this information has been already reported in the literature. Such images and their corresponding HOMFLY polynomial will be unclassified. Using 52 vertices, we could classify all fractures. This was not the case when we increased the number of coordinates to 108 by using additional kernels in Table \ref{tab:filter}. The precision was computed as described in Training and Validation. For the class representing the ribs without fractures we obtained a precision of 0.52 and with fractures a precision of 0.60.

\section{Discussion}
In this study we explored the possibility to combine classical image processing with topology, a branch of mathematics dealing with the classification of objects having variable size, shape, and orientation. Current deep learning architectures are based on multi-layer artificial neural networks. These models can be trained to increase their accuracy. However, the whole deep learning method is based on a statistical approach \cite{krizhevsky2012,schmidhuber2015}. On the contrary, a topological invariant such as the polynomial knot invariant has an “infinite accuracy”. To introduce a polynomial knot invariant into a deep learning process the image has to be converted into a topological object such as a closed curve. Our strategy consisted of converting the pixel values into a set of three-dimensional coordinates and building a closed curve in a three-dimensional Euclidian space by joining all the coordinates. We assumed that the variations in size, color and shape contained in the images showing the same object will translate into small geometrical variations of the computed closed curve. As long as the geometrical variations are small enough, two distinct objects will produce two distinct curves with different topological properties. By computing the polynomial knot invariant of the curve representing an object, one can uniquely classify it. The original problem of LeCun \textit{et al.} \cite{lecun1985} is actually very similar to what topology tends to achieve, namely the classification of mathematical objects that can be distorted indefinitely without changing their properties. It is therefore very tempting to borrow techniques from topology and make them compatible with deep learning.

We classified closed curves embedded in three-dimensional Euclidean space using the HOMFLY polynomial knot invariant. Each axis of the three-dimensional Euclidean space represents one the RGB channel and we applied a color mapping technique to convert the grayscale values of the CT image into an RGB image. By applying a specific kernel function on CT images representing the same object, one would expect to produce similar coordinate values that can be isolated in a region in the three-dimensional Euclidean space represented by the RGB color space. The distance between each region for a specific kernel function or cluster of coordinates should be large enough such that when the centroids of the clusters are joined to produce a piecewise-linear curve, the knot type obtained through this process should remain the same if new objects are analyzed and their coordinates joined in the same manner. In other words, a new image should produce a set of coordinates that match the actual cluster. Hence, the classes should populate different regions such that when the centroids are joined, each class produces a different knot type. For medical images our approach turned out to be more complicated as we ended up with more than one knot type for a given class. Therefore, we modified our strategy and, instead of computing the knot type obtained by joining the centroid of each cluster of coordinates, we computed the polynomial of each image and listed them separately: one for the image showing a rib fracture and one for those without a fracture. The precision was computed as the fraction of polynomials that were present in only one list. One major limitation by adapting the original idea to rib fracture classification is the number of different polynomials present during the analysis. Our current dataset is not large enough to encompass all possible polynomials. Hence, although we can show this approach can classify rib fracture with a precision of 0.60, we could not use this pipeline for predicting unlabelled images. Furthermore, it is not excluded that a larger dataset can substantially improve our current precision.
It is also worth noting that the two models based on Tensorflow give comparable results. Whether we used the MobileNet version 1.0 architecture comprising a large number of layers or a simplified architecture with three convolutional and two dense layers, the results did not differ substantially. MobileNet has been trained on ImageNet, a large database of images found on Internet. These images represent everyday objects which have in general distinct contours or relatively sharp borders. Such images seem adequate for the type of kernels used commonly in deep learning. Medical images do not necessarily display sharp contours and therefore may not be suited for the current set of kernel functions \cite{cao2017}. Which set of kernels can better abstract information in medical images and more specifically rib fractures might need further investigation.

Currently, our approach still suffers from producing multiple topologies from the same objects albeit the topologies from one object do not completely overlap with the topologies produced by another object. This allow us to classify two types of objects, in our case the medical findings, into two categories by having a list of topologies characterising one and the other. However, the ultimate goal is not yet achieved and the preprocessing of the images using the kernel techniques is not yet satisfactory. We believe we can find the appropriate approach by looking at other types of image analyses while thinking of other ways of transforming the pixel values into coordinates.
Having such an algorithm for medical images can bring the field forward. It will allow to have a much greater accuracy during the analysis, which is instrumental for a proper diagnosis.

\section{Conclusion}
Modern computed tomography scanners generate huge amounts of image data and in the case of forensic postmortem investigations a dataset can have more than thousands of single images. Hence, there is a demand for automation. Further progress is needed as the workload for forensic pathologists has increased. Here, we combined classical image analysis using kernel functions with topology to devise a machine learning algorithm for medical image classification. Although the current method requires further optimisation, we have shown how it can be applied to classify CT images of rib fractures and potentially other types of medical images.

\section{Acknowledgements}
The authors at the Zurich Institute of Forensic Medicine express their gratitude to the Emma Louise Kessler Foundation for supporting this research. This material is based upon work supported by the National Science Foundation under Grant No.~1720342 to EJR. The authors declare having no conflicting interests.

\section{Supplementary Materials}

\begin{figure}
\begin{center}
\includegraphics[keepaspectratio = true, height=4.0in]{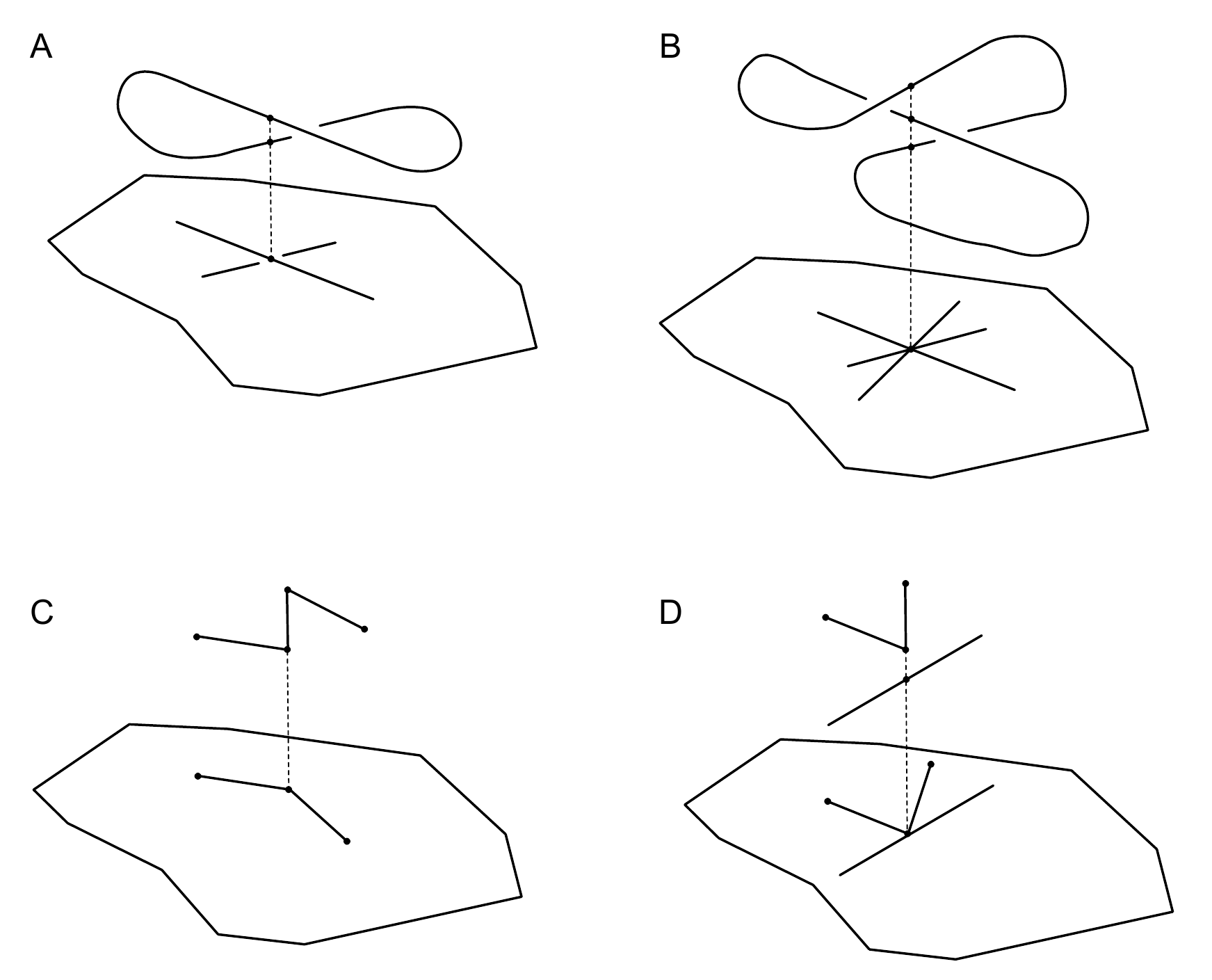}
\caption{Crossing convention for projecting a knot onto a 2D plane. (A) Regular projection of a knot. (B to D) Degenerate projections of a knot. In (B) the projection is degenerate due to triple crossings. In (C) one edge is normal to the projection plane preventing from visualizing possible crossings and reduces the projected edge to a point. In (D) the orientation of the projection is preventing from deciding which part of the knot lies above and which part below.}
\label{sup1}
\end{center}
\end{figure}

\begin{figure}
\begin{center}
\includegraphics[keepaspectratio = true, height=4.6in]{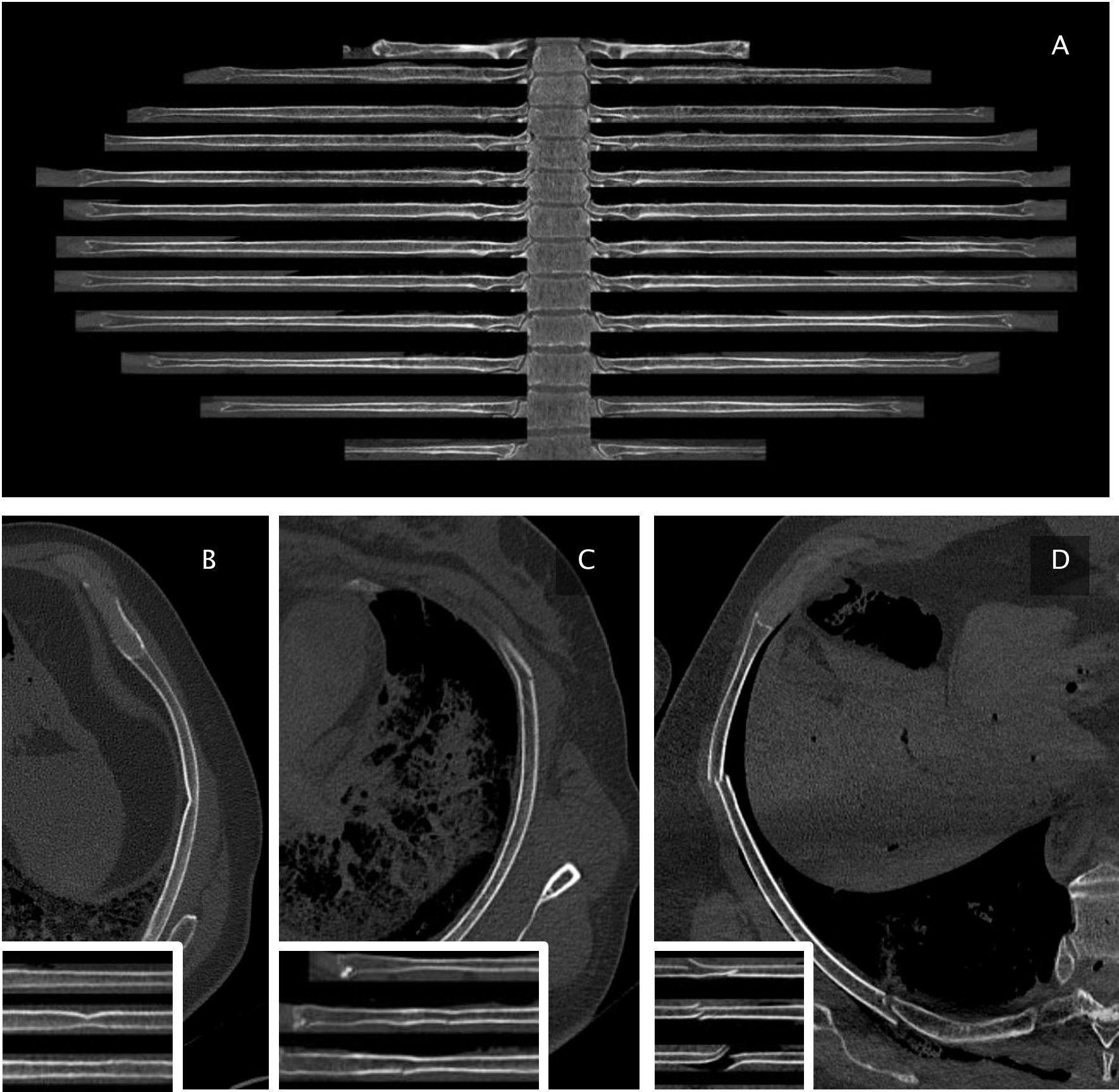}
\caption{Rib unfolding. (A) Rib unfolding with no rib fractures and 12 ribs on either side of the vertebral column. (B) Multi-planar view of a so-called buckle rib fracture (subgroup of incomplete rib fractures). Inset lower left: Rib unfolding view of a buckle rib fracture with no interrupted cortical line but obvious kinking. (C) Multi-planar view of an incomplete rib fracture; outer cortical line interrupted. Inset lower left: Rib unfolding view of an incomplete rib fracture with interrupted cortical line. (D) Multi-planar view of two complete rib fractures in the middle and the back of the rib. In contrast to the other fractures, both cortical lines are interrupted in complete fractures. Inset lower left: Rib unfolding view of 3 complete rib fractures.}
\label{sup2}
\end{center}
\end{figure}

\end{document}